\documentclass[conference]{IEEEtran}
\IEEEoverridecommandlockouts
\usepackage{cite}
\usepackage{amsmath,amssymb,amsfonts}
\usepackage{braket}
\usepackage{algorithmic}
\usepackage{graphicx}
\usepackage{textcomp}
\usepackage{xcolor}
\usepackage{adjustbox}

\def\BibTeX{{\rm B\kern-.05em{\sc i\kern-.025em b}\kern-.08em
    T\kern-.1667em\lower.7ex\hbox{E}\kern-.125emX}}
    
\usepackage{balance}
    
\begin{document}

\title{Application of Power Flow problem to an open quantum neural hardware\\

\thanks{??????????????????/\$31.00  \copyright 2023 IEEE}
}

\author{\IEEEauthorblockN{1\textsuperscript{st}Ekin Erdem Aygül }
\IEEEauthorblockA{\textit{Department of Electrical Engineering} \\
\textit{{I}stanbul Technical University}\\
{I}stanbul, Turkey \\
aygule16@itu.edu.tr}
\and
\IEEEauthorblockN{2\textsuperscript{nd} Melih Can Topal}
\IEEEauthorblockA{\textit{Department of Electrical Engineering} \\
\textit{{I}stanbul Technical University}\\
{I}stanbul, Turkey \\
topal17@itu.edu.tr}
\and
\IEEEauthorblockN{3\textsuperscript{rd} Ufuk Korkmaz}
\IEEEauthorblockA{\textit{Department of Electrical Engineering} \\
\textit{{I}stanbul Technical University}\\
{I}stanbul, Turkey \\
ufukkorkmaz@itu.edu.tr}
\and

\IEEEauthorblockN{4\textsuperscript{th} Deniz T\"{u}rkpen\c{c}e}
\IEEEauthorblockA{\textit{Department of Electrical Engineering} \\
\textit{{I}stanbul Technical University}\\
{I}stanbul, Turkey \\
dturkpence@itu.edu.tr}

}

\maketitle

\begin{abstract}
Significant progress in the construction of physical hardware for quantum computers has necessitated the development of new algorithms or protocols for the application of real-world problems on quantum computers. One of these problems is the power flow problem, which helps us understand the generation, distribution, and consumption of electricity in a system. In this study, the solution of a balanced 4-bus power system supported by the Newton-Raphson method is investigated using a newly developed dissipative quantum neural network hardware. This study presents the findings on how the proposed quantum network can be applied to the relevant problem and how the solution performance varies depending on the network parameters.
\end{abstract}


\section{Introduction}

Various mathematical solution proposals have been introduced for the Power Flow (PF)~\cite{grainger1994,stott1974,balamurugan2011,shirmohammadi1988,chen1991,tiwari2018,zhou2018,pham2022}, which is an engineering problem related to the distribution of electric energy, still remaining an indispensable energy source in our modern world. The most conventional and fundamental techniques employed with the YBUS matrix encompass Gauss-Seidel power flow, Newton-Raphson (NR) power flow, and Decoupled methods~\cite{grainger1994}. YBUS methods offer a reliable and efficient solution for a wide range of power systems. 

However, these methods have certain limitations. The Newton-Raphson (NR) method incurs high computational costs due to the need for calculating the Jacobian matrix. On the other hand, the Decoupled method, while computationally less expensive, exhibits a higher margin of error~\cite{stott1974}. In addition, there exist alternative methods tailored to different power system configurations. For instance, the forward and backward sweep method is employed for radial networks~\cite{balamurugan2011}. In the case of loosely interconnected power systems, a compensation method has been developed~\cite{shirmohammadi1988}. Moreover, a novel approach utilizing ZBUS and LU triangularization has been proposed for both balanced and unbalanced power systems~\cite{chen1991}. There are numerous variations of the aforementioned methods. For example, a study was conducted on the IEEE 14-bus system, comparing different optimizers that take load reactive and active powers as inputs, with voltage magnitudes being returned as outputs~\cite{tiwari2018}.

Artificial neural networks, which are learning systems, constitute another method among those mentioned above. For instance, InterPSS was utilized for generating the training dataset~\cite{zhou2018}. Pham and Li, on the other hand, compared their ANN studies using the ReLU activation function with the DC power flow method in terms of accuracy and speed~\cite{pham2022}. Another study involved creating a dataset using 21 base cases from the Saudi national grid to train an ANN, and the results were then compared with the Newton-Raphson method ~\cite{alsulami2017}.

In this study, we propose a solution to the power flow (PF) problem by developing a quantum neural network that operates using a dissipative quantum hardware, in which its building block has recently been introduced~\cite{korkmaz_quantum_2023,korkmaz2022}. To accomplish this, we employ the standard theory of artificial neural learning~\cite{dasilva2016} and adapt our current problem to the hardware parameters that have been introduced. 

Although quantum neural networks do not possess proven explicit superiority over their classical counterparts, their widespread use stems from incorporating quantum subroutines and thereby increasing the potential for quantum advantage in various problem-solving applications. In previous work, it was demonstrated that dissipative quantum computation, which is emphasized in this study, is equivalent to the standard quantum circuit model~\cite{verstraete_quantum_2009}.
\section{Problem Definition}
\subsection{Newton Raphson Power Flow}
Newton-Raphson method is an iterative method which takes advantage of Taylor series expansion and first-order approximation ~\cite{grainger1994}. We can express the general definition as 
\begin{equation}
\begin{aligned}
&g_1\left(x_1, x_2, u\right)=h_1\left(x_1, x_2, u\right)-b_1=0\\
&g_2\left(x_1, x_2, u\right)=h_2\left(x_1, x_2, u\right)-b_2=0
\end{aligned}
\end{equation}
and the  Taylor series expansion of these multivariable functions can be written as \begin{equation}
\begin{aligned}
g_1\left(x_1^*, x_2^*, u\right)= & g_1\left(x_1^{(0)}, x_2^{(0)}, u\right) \\
& +\left.\Delta x_1^{(0)} \frac{\partial g_1}{\partial x_1}\right|^{(0)} +\left.\Delta x_2^{(0)} \frac{\partial g_1}{\partial x_2}\right|^{(0)}+\cdots \\
g_2\left(x_1^*, x_2^*, u\right)= & g_2\left(x_1^{(0)}, x_2^{(0)}, u\right) \\
& +\left.\Delta x_1^{(0)} \frac{\partial g_2}{\partial x_1}\right|^{(0)} +\left.\Delta x_2^{(0)} \frac{\partial g_2}{\partial x_2}\right|^{(0)}+\cdots
\end{aligned}
\end{equation}
Since the exact solution of the set of functions $g_1\left(x_1^*, x_2^*, u\right)$ and $g_2\left(x_1^*, x_2^*, u\right)$is equal to zero, neglecting terms equal or higher than second order differential the Taylor series expansion of this set can be written like
\begin{equation}
\left[\begin{array}{ll}
\frac{\partial g_1}{\partial x_1} & \frac{\partial g_1}{\partial x_2} \\
\frac{\partial g_2}{\partial x_1} & \frac{\partial g_2}{\partial x_2}
\end{array}\right]\left[\begin{array}{l}
\Delta x_1^{(0)} \\
\Delta x_2^{(0)}
\end{array}\right]=\left[\begin{array}{l}
0-g_1\left(x_1^{(0)}, x_2^{(0)}, u\right) \\
0-g_2\left(x_1^{(0)}, x_2^{(0)}, u\right)
\end{array}\right]
\end{equation}
If we rewrite the equation in the simpler form 
\begin{equation}
J^{(0)}\left[\begin{array}{l}
\Delta x_1^{(0)} \\
\Delta x_2^{(0)}
\end{array}\right]=\left[\begin{array}{l}
\Delta g_1^{(0)} \\
\Delta g_2^{(0)}
\end{array}\right]
\end{equation}
Here, the matrix $J$ is the Jacobian matrix. After the algorithm starts initial values of $x^{(0)}$, the jacobian is formed. By taking the inverse of Jacobian, mismatches are found. From mismatch equation new values of $x$ are found using $x_i^{(k+1)}=x_i^{(k)}+\Delta x_i^{(k)}$. This process is repeated until mismatch is close to zero or under a specified error rate.
\subsection{Power Flow} 
Power networks are composed of buses that elements in the power network are connected and lines that connect these buses. YBUS methods can be used to define Power Flow Problem ~\cite{dasilva2016}.
\begin{equation}
\begin{gathered}
P_i=\left|V_i\right|^2 G_{i i}+\sum_{\substack{n=1 \\
n \neq i}}^N\left|V_i V_n Y_{i n}\right| \cos \left(\Theta_{i n}+\delta_n-\delta_i\right) \\
Q_i=-\left|V_i\right|^2 B_{i i}-\sum_{\substack{n=1 \\
n \neq i}}^N\left|V_i V_n Y_{i n}\right| \sin \left(\Theta_{i n}+\delta_n-\delta_i\right)
\end{gathered}
\end{equation}
where power mismatch is defined as $\Delta P_i=P_{i, s c h}-P_{i, c a l c}$ and $\Delta Q_i=Q_{i, s c h}-Q_{i, c a l c}$. The generalization of the linear matrix system presented below 
\begin{equation}
[J]\left[\begin{array}{c}
\Delta \delta \\
\frac{\Delta|V|}{|V|}
\end{array}\right]=\left[\begin{array}{l}
\Delta P \\
\Delta Q
\end{array}\right].
\end{equation}
is derived from the application of Taylor series. In the formation of this linear system of equations, the slack bus is excluded. Furthermore, for PV buses, voltage corrections are consistently set to zero, and their reactive power is left unspecified. Consequently, the columns multiplied by zero and the rows with an indeterminate solution are also excluded. As for PQ buses, no omissions are made; however, since the voltage magnitude and phase are unknown for these buses, they are estimated. Considering Eq. (15),  the elements of Jacobian can be divided into four matrices $J_{11}, J_{12}, J_{21}, J_{22}$ where elements of each one of these matrices are $\frac{\partial P_i}{\partial \delta_j},\left|V_j\right| \frac{\partial P_i}{\partial\left|V_j\right|}$, $\frac{\partial Q_i}{\partial \delta_j}$, $\left|V_j\right| \frac{\partial Q_i}{\partial\left|V_j\right|}$ respectively and i and j represent the row and column number.
To form the Jacobian these elements must be calculated. Using Eq.~(\ref{Eq:v new}) and Eq.~(\ref{Eq:delta new}), the formulas below are obtained.
\begin{align}
\left|V_j\right| \frac{\partial Q_i}{\partial\left|V_j\right|}= -\left|V_i V_j Y_{i j}\right| \sin \left(\theta_{i j}+\delta_j-\delta_i\right)=M_{i j} \\
-\left|V_j\right| \frac{\partial P_i}{\partial\left|V_j\right|}= -\left|V_i V_j Y_{i j}\right| \cos \left(\theta_{i j}+\delta_j-\delta_i\right)=N_{i j}
\end{align}

And for \textit{i}=\textit{j}, 
\begin{equation}
\begin{aligned}
&\frac{\partial P_i}{\partial \delta_i}=-\sum_{\substack{n=1 \\ n \neq i}}^N M_{i n}=M_{i i}\\
&\begin{gathered}
\left|V_i\right| \frac{\partial Q_i}{\partial\left|V_i\right|}=-M_{i i}-2\left|V_i\right|^2 B_{i i}\\
\end{gathered}
\end{aligned}
\end{equation}

\begin{equation}
\begin{gathered}
\frac{\partial Q_i}{\partial \delta_i}=-\sum_{\substack{n=1 \\
n \neq i}}^N N_{i n}=N_{i i} \\
\left|V_i\right| \frac{\partial P_i}{\partial\left|V_i\right|}=N_{i i}+2\left|V_i\right|^2 G_{i i}
\end{gathered}
\end{equation}
next, the obtained Jacobian inverse is used to find mismatches of voltage magnitudes and phases. The updated parameters are given as
\begin{equation}\label{Eq:v new}
\left|V_i\right|^{\text {new }}=\left|V_i\right|^{\text {old }}\left(1+\frac{\Delta\left|V_i\right|^{\text {old }}}{\left|V_i\right|^{\text {old }}}\right),
\end{equation}

\begin{equation}\label{Eq:delta new}
\delta^{n e w}=\delta^{o l d}+\Delta \delta^{o l d}.
\end{equation}
\subsection{Artificial Neural Network}
Hodgkin and Huxley conducted groundbreaking research on the electrical properties of biological neurons where they were able to describe their electrical behavior~\cite{hodgkin1952}. The ability to mathematically describe biological material in this way has led to the idea that brain nerve cells can be represented in this manner. Inspired by the firing behavior of brain cells, Rosenblatt introduced a mathematical binary classifier model called a perceptron ~\cite{mcculloch1943}, which is represented by the following mathematical expression
\begin{equation}
\begin{aligned}
& I=\sum_{i=1}^n x_i W_i \\
& Y=g(I)
\end{aligned}
\end{equation}
where, $x_i$ are inputs, $W_i$ are weights, and $g$ is the activation function.

The perceptron utilizes Hebb's learning rule during its learning procedure~\cite{morris1999}. This rule dictates that the weights of the neuron are adjusted when an error occurs between the desired output and the actual output. Conversely, if there is no discrepancy, the weights remain unaltered. The iteration procedure is given as
 
\begin{equation}
\boldsymbol{W}_i^{(\text {current })}=\boldsymbol{W}_i^{(\text {previous })}+\eta\left(d^{(k)}-y^{(k)}\right) \boldsymbol{x}^{(k)}
\end{equation}
where, $k$ is the number determining which sample is used,  $\boldsymbol{W}_i$ is a vector containing the weights, $d^{(k)}$ is the desired output for the $\mathbf{k}^{\text {th }}$ sample, $\boldsymbol{x}^{(k)}$ is the input vector for the $\mathbf{k}^{\text {th }}$ sample and $\eta $ is the learning rate. When it comes to training a multilayer neural network, a new notation and a re-expression of the error function minimization in terms of the structure of the multilayer network are required.

\begin{figure}[ht]%[htbp]
\centerline{\includegraphics[width=3.0in]{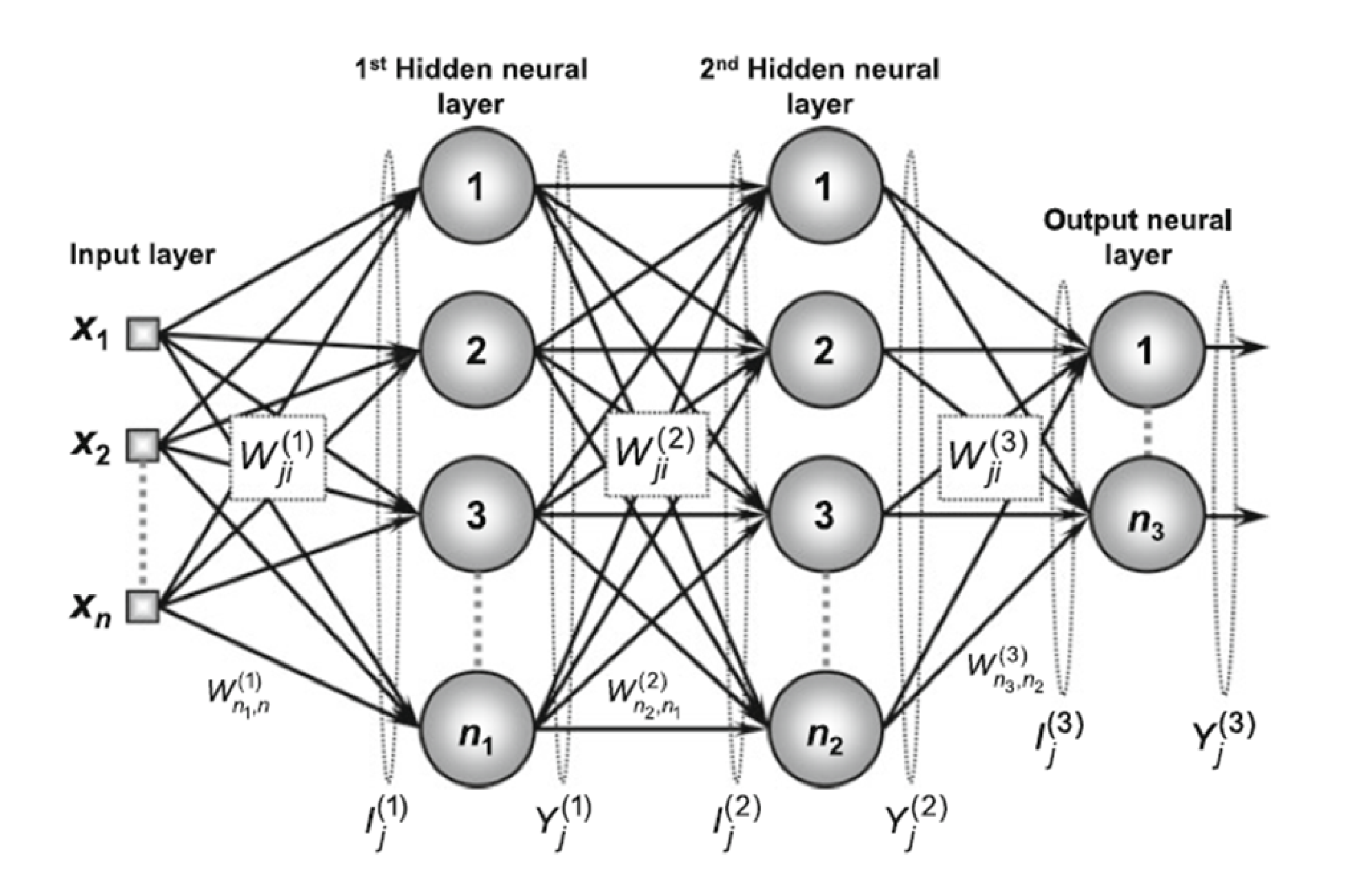}}
\caption{Multi Layer Feed Forward Neural Network Scheme with Notations ~\cite{dasilva2016}}
\label{fig:fig1}
\end{figure}

In Fig.~\ref{fig:fig1}, $Y_i^{(L)}$ output of the $i^{\text {th }}$ neuron in layer $L$, $I_i^{(L)}$ is the Input to activation function of $i^{\text {th }}$ neuron in layer $L$, $W_{j i}^{(L)}$ is the Weight of the connection between $j^{\text {th }}$ neuron in layer $L$ and $i^{\text {th }}$ neuron in layer $(L-1)$, $g(.)$ is an activation function, $n_i$ is the number of neurons in the $i^{\text {th }}$ layer. In this architecture, each neuron is densely connected to every neuron in the preceding and subsequent layers. This connectivity pattern is referred to as dense connections. The formulas for the defined variables are provided below.
\begin{equation}
\begin{aligned}
& I_i^{(L)}=\sum_{i=1}^{n_{L-1}} Y_i^{(L-1)} W_{j i}^{(L)} \\
& Y_i^{(L)}=g\left(I_i^{(L)}\right)
\label{eq:eq15}
\end{aligned}
\end{equation}
\subsubsection{Backpropagation Algorithm}

Backpropagation algorithm uses gradient descent method. The gradient of the error with respect to $\mathrm{W}_{\mathrm{ji}}$ gives can be written by expanding the formula using chain rule.
\begin{equation}
\nabla E^{(L)}=\frac{\partial E}{\partial W_{j i}^{(L)}}=\frac{\partial E}{\partial Y_j^{(L)}} \frac{\partial Y_j^{(L)}}{\partial I_j^{(L)}} \frac{\partial I_j^{(L)}}{\partial W_{j i}^{(L)}}
\end{equation}
Here for the output layer with mean square error loss function.
\begin{equation}
\begin{aligned}
&\frac{\partial I_j^{(L)}}{\partial W_{j i}^{(L)}}=\partial Y_j^{(L-1)};
&\begin{aligned}
& \frac{\partial Y_j^{(L)}}{\partial I_j^{(L)}}=g\left(I_j^{(L)}\right) \\
& \frac{\partial E}{\partial Y_j^{(L)}}=-\left(d_j-Y_j^{(L)}\right)
\end{aligned}
\end{aligned}
\end{equation}
We define the delta as 
\begin{equation}
\delta_j^{(L)}=\left(d_j-Y_j^{(L)}\right) g\left(I_j^{(L)}\right)=-\frac{\partial E}{\partial Y_j^{(L)}} \frac{\partial Y_j^{(L)}}{\partial I_j^{(L)}}
\end{equation}

But for the hidden layers it is harder to find as $-\frac{\partial E}{\partial Y_j^{(L)}}$.

\begin{equation}
\frac{\partial E}{\partial Y_j^{(L-1)}}=\sum_{k=1}^{n_L} \frac{\partial E}{\partial I_k^{(L)}} \frac{\partial\left(\sum_{k=1}^{n_L} W_{k j}^{(L)} Y_j^{(L-1)}\right)}{\partial Y_j^{(L-1)}}
\end{equation}
Above in 17, it can be seen that $\frac{\theta E}{\Delta I_k^{(L)}}$ is actually the $\delta_j^{(L)}$ of the layer ahead. As we calculate these $delta$ values every time they will be stored to calculate the $delta$ of the previous layer. 
\begin{equation}
\frac{\partial E}{\partial Y_j^{(L-1)}}=-\sum_{k=1}^{n_L} \delta_j^{(L)} W_{k j}^{(L)}
\end{equation}
Now we can calculate the delta for the $\left(\mathrm{L}-\underline{1}^{\mathrm{th}}\right.$ layer.
\begin{equation}
\delta_j^{(L)}=\left(\sum_{k=1}^{n_L} \delta_j^{(L)} W_{k j}^{(L)}\right) g\left(I_j^{(L-1)}\right)
\end{equation}
A generalized version of updating the weight is given as 
\begin{equation}
W_{j i}^{(L)}(t+1)=W_{j i}^{(L)}(t)+\eta \delta_j^{(L)} Y_i^{(L-1)}
\end{equation}
Here $\eta$ is the learning rate. The learning rate does not have to be fixed. An adaptive learning rate can be beneficial which will be mentioned in the following. 
Here since this is plain back propagation, all of the weights are updated at the same time after the error and the gradient for that error is found for every sample. 
\section{Quantum Neural Network}
In the open quantum counterpart of the Artificial Neural Network (ANN) depicted in Figure 1, the nodes are substituted with quantum spins characterized by spin numbers $J \geq 1/2$, while the input layer is substituted with reservoirs carrying quantum information. The operational mechanism of the introduced quantum neural network relies on a repeated interaction process and Completely Positive and Trace-Preserving (CPTP) maps.

\subsection{Open Quantum Systems}

Assuming a system with the density matrix $\rho$ and the environment denoted with $\rho_E$ are in a product state. Even though the system is not unitary, total of system and environment is unitary. Therefore, for an arbitrary unitary transformation $U$ the evaluation of the system+environment is given as
\begin{equation}
\rho_{s y s} \otimes \rho_{e n v} \rightarrow U \rho_{s y s} \otimes \rho_{e n v} U^{\dagger}.
\end{equation}
Next, the system of interest is obtained by a partial trace operation which is a non-unitary evolution  
\begin{equation}
\rho_{s y s}=\text{Tr}_{\text{env}}\left(\rho_{s y s+e n v}\right).
\end{equation}

In order to describe the evolution of a state in open quantum environment, we used quantum dynamical map. In general quantum dynamical map can be defined as 
\begin{equation}
\rho_{s y s}^{\prime}=\varepsilon(\rho)=\text{Tr}_{\text{env}}\left(U \rho_{s y s+e n v} U^{\dagger}\right).
\end{equation}
There are several conditions that required for the defining physical process using dynamical maps. Firstly a quantum map must preserve unit trace $\operatorname{Tr}(\varepsilon(\rho))=\operatorname{Tr}(\rho)=1$. Secondly a quantum map must be convex linear. $\varepsilon\left(\sum_i p_i \rho_i\right)=\sum_i p_i \varepsilon\left(\rho_i\right)$
And lastly the dynamical map must be completely positive~\cite{korkmaz_quantum_2023}.
On top of that, if a weak coupling condition is met where the additivity of the quantum dynamical map is valid, a system can be written in a linear combination of dynamical maps ~\cite{kolodynski2018}.
\begin{equation}
\Lambda\left(\rho_0\right)=\sum_i P_i \Phi^{(i)}\left(\rho_0\right)
\label{eq:eq26}
\end{equation}
\subsection{Multi-layer Quantum Neural Network}

Designing a dissipative quantum neural network composed of perceptrons operating based on the aforementioned principles is a straightforward task. It is important to highlight that the hardware implementation relies on the weak coupling condition, allowing for the dissipative transfer of quantum data from the pure information reservoirs to the network depicted below ~\cite{korkmaz_quantum_2023,korkmaz2022b}
\begin{equation}\label{Eq:Rho_tp}
\rho_{R_i}=\bigotimes_{k=1}^n \rho_k\left(\theta_i, \phi_i\right).
\end{equation}
As stated in Eq.~(\ref{Eq:Rho_tp}) $\theta$ and $\phi$ are used to define a pure quantum state in the Bloch sphere. Information reservoirs are composed of many quantum states of the same state.
The dynamical map that defines the interaction between the system and ith information reservoir is given as
\begin{equation}
\begin{aligned}
\Phi_{n \tau}^{(i)} =&\operatorname{tr}_n\left(U_{0_{i n}} \ldots t r_1\left(U_{0_{i 1}}\left(\rho_0 \otimes \rho_{R_{i 1}}\right) U_{0_{i 1}}^{\dagger}\right) \otimes \ldots\right. \\
& \left.\ldots \otimes \rho_{R_{i n}} U_{0_{i n}}^{\dagger}\right)
\end{aligned}
\end{equation}
where, $U_{0_{i k}}=e^{-i H_{0 i}^k \tau}$. 

In this context, the variable $n\tau$ represents the time required for $n$ collisions. With a finite number of collisions, it is assumed that the probe qubit can attain a steady state~\cite{bruneau2014}. To derive the unitary propagator, a master equation is employed, which is based on a micromaser-like repeated interactions approach~\cite{korkmaz_quantum_2023,korkmaz2022b,korkmaz2021a,korkmaz2022} given below
\begin{equation}
\begin{aligned}
U(\tau)=&1-i \tau(\left.\sigma_0^{+} J_{g i}^{-}+\sigma_0^{-} J_{g i}^{+}\right) \\
& -\frac{\tau^2}{2}\left(\sigma_0^{+} \sigma_0^{-} J_{g i}^{-} J_{g i}^{+}+\sigma_0^{-} \sigma_0^{+} J_{g i}^{+} J_{g i}^{-}\right).
\end{aligned}
\end{equation}
The steady-state density matrix of the probe qubit is obtained as~\cite{korkmaz_quantum_2023}
\begin{equation}
\begin{aligned}
\rho_0^{s s}=& \frac{1}{\sum_i^N g_i{ }^2} \sum_{i=1}^N g_i{ }^2\left(\left\langle\sigma_i^{+} \sigma_i^{-}\right\rangle|e\rangle\left\langle e\left|+\left\langle\sigma_i^{-} \sigma_i^{+}\right\rangle\right| g\right\rangle\langle g|\right. \\
&+i \gamma_1^{-}\left(\left\langle\sigma_i^{+} \sigma_i^{-}\right\rangle-\left\langle\sigma_i^{-} \sigma_i^{+}\right\rangle\right)|e\rangle\langle g|
+\text { H.c. })
\end{aligned}
\end{equation}
In this context, $\left\langle\sigma_z^0\right\rangle^{s s}$ corresponds to the outputs of the artificial neurons in an artificial neural network (ANN), while $\left\langle\sigma_z\right\rangle_i$ can be interpreted as the outputs of the preceding neurons. Therefore, the formula mentioned above incorporates Eq.~(\ref{eq:eq15}). Additionally, the weights in the ANN are represented by the coupling strength $g$. 

\subsection{Activation Function}

Activation functions play a vital role in ANNs. However, the specific activation function used in the proposed QNN remains unspecified. Graphical methods are employed to determine the appropriate activation function for the system. Additionally, it has been shown that adjusting the spin number influences the steepness of the resulting hyperbolic tangent function. 

\begin{figure}[ht]%[htbp]
\centerline{\includegraphics[width=3.0in]{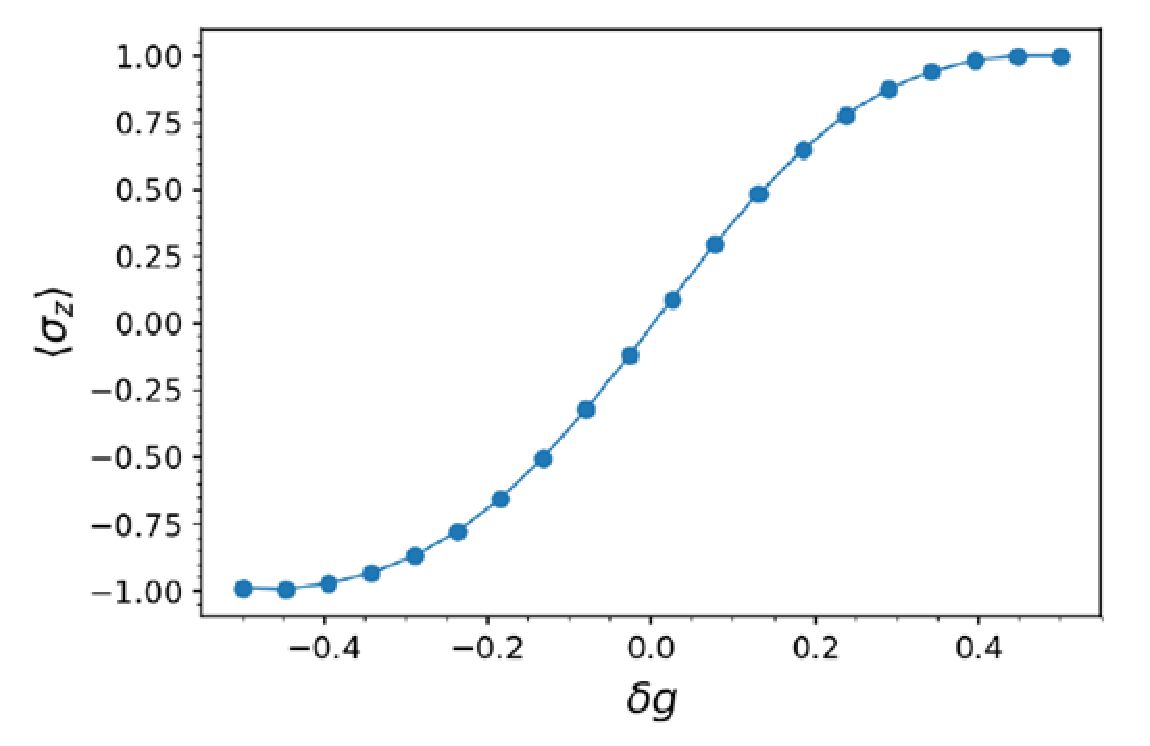}}
\caption{Obtained activation function ~\cite{korkmaz2022}. The probe qubit,initially in the $\ket{+}=(\ket{e}+\ket{g})/\sqrt{2}$  state, made collisional contact with identical reservoir units, $\ket{\Psi(\theta,\phi)}$ , which had $\theta=0$, $\phi=0$ and $\theta=\pi$, $\phi=0$. $\Gamma=2\times10^{-5}$ is the decay rate of the probe qubit. The reservoir coupling strength is $g=0.01$. The interaction time between the probe qubit and the reservoir is $\tau=3$.
These parameters are dimensionless and scaled by superconducting resonator frequency $\omega_r$}
\label{fig:fig2}
\end{figure}

Steady state qubit magnetization is the merit quantifier of the quantum neurons as output data. The expectation value is obtained by measuring in the $z$-axis giving (Fig.~\ref{fig:fig2})
\begin{equation}
\left\langle\sigma_z^0\right\rangle^{s s}=\frac{1}{\sum_i^N g_i^2} \sum_i^N g_i{ }^2\left\langle\sigma_z\right\rangle_i.
\end{equation}

\begin{figure}[ht]%[htbp]
\centerline{\includegraphics[width=3.0in]{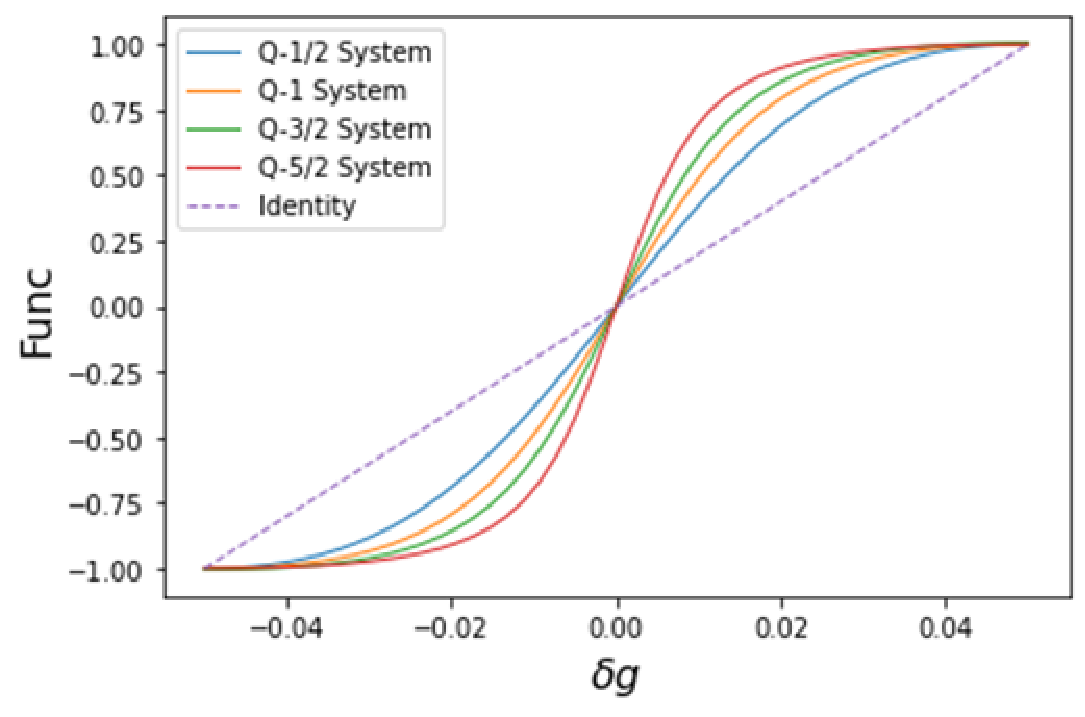}}
\caption{Change of steepness of tanh based on spin number. }
\label{fig:fig3}
\end{figure}
Using curve fitting techniques, the activation functions are found as dependent on a new variable $\beta$ which depends on spin number (Fig.~\ref{fig:fig3}).

\begin{equation}
g(I')=\tanh (\beta I)
\end{equation}
As we can see from the equation, different spin numbers refer to different beta values. For spin numbers $1/2, 1, 3/2, 5/2$, we get $\beta$ values $2.22, 2.78, 3.33, 4.1$ respectively.

\section{Simulation}
\subsection{Power System}
We chose a simple 4-bus network for simulations in Table~1~\cite{chen1991}. YBUS data and other specifications are given in Table~2.

\begin{table}[]
\centering
\begin{adjustbox}{width=0.5\textwidth}
\begin{tabular}{lccccc}
\multicolumn{6}{l}{Table 1: Power Network Bus Data }                         \\ \cline{2-5}
\multicolumn{1}{l|}{}     & \multicolumn{2}{c|}{Generation} & \multicolumn{2}{c|}{Load} & \multicolumn{1}{l}{}    \\ \hline
\multicolumn{1}{|l|}{Bus} &
  \multicolumn{1}{l|}{\textit{P, MW}} &
  \multicolumn{1}{l|}{\textit{Q, Mvar}} &
  \multicolumn{1}{l|}{P, MW} &
  \multicolumn{1}{l|}{Q, Mvar} &
  \multicolumn{1}{l|}{V, per unit} \\ \hline
\multicolumn{1}{c}{1}     & -                & -            & 50         & 30.99        & 1.00 / 0$^{\circ}$                \\
2                         & 0                & 0            & 170        & 105.35       & 1.00 / 0$^{\circ}$               \\
3                         & 0                & 0            & 200        & 123.94       & 1.00 / 0$^{\circ}$                \\
4                         & 318              & -            & 80         & 49.58        & 1.02 / 0$^{\circ}$               
\end{tabular}
\label{tab:tablo1}
\end{adjustbox}
\end{table}

\begin{table}[]
\centering
\begin{adjustbox}{width=0.5\textwidth}
\begin{tabular}{lllll}
\multicolumn{5}{l}{Table 2: Power Network YBUS Data} \\ \hline
\multicolumn{1}{|l|}{Bus no.} &
  \multicolumn{1}{l|}{(1)} &
  \multicolumn{1}{l|}{(2)} &
  \multicolumn{1}{l|}{(3)} &
  \multicolumn{1}{l|}{(4)} \\ \hline
\multicolumn{1}{|l|}{(1)} &
  \multicolumn{1}{l|}{\begin{tabular}[c]{@{}l@{}}8.985190\\ -j 44.835953\end{tabular}} &
  \multicolumn{1}{l|}{\begin{tabular}[c]{@{}l@{}}-3.815629\\ +j 19.078144\end{tabular}} &
  \multicolumn{1}{l|}{\begin{tabular}[c]{@{}l@{}}-5.169561\\ +j 25.847809\end{tabular}} &
  \multicolumn{1}{l|}{0} \\ \hline
\multicolumn{1}{|l|}{(2)} &
  \multicolumn{1}{l|}{\begin{tabular}[c]{@{}l@{}}-3.815629\\ +j 19.078144\end{tabular}} &
  \multicolumn{1}{l|}{\begin{tabular}[c]{@{}l@{}}8.985190 \\ -j 44.835953\end{tabular}} &
  \multicolumn{1}{l|}{0} &
  \multicolumn{1}{l|}{\begin{tabular}[c]{@{}l@{}}-5.169561 \\+j 25.847809\end{tabular}} \\ \hline
\multicolumn{1}{|l|}{(3)} &
  \multicolumn{1}{l|}{\begin{tabular}[c]{@{}l@{}}-5.169561 \\+j 25.847809\end{tabular}} &
  \multicolumn{1}{l|}{0} &
  \multicolumn{1}{l|}{\begin{tabular}[c]{@{}l@{}}8.193267 \\-j 40.863838\end{tabular}} &
  \multicolumn{1}{l|}{\begin{tabular}[c]{@{}l@{}}-3.023705 \\+j 15.118528\end{tabular}} \\ \hline
\multicolumn{1}{|l|}{(4)} &
  \multicolumn{1}{l|}{0} &
  \multicolumn{1}{l|}{\begin{tabular}[c]{@{}l@{}}-5.169561 \\+j 25.847809\end{tabular}} &
  \multicolumn{1}{l|}{\begin{tabular}[c]{@{}l@{}}-3.023705 \\+j 15.118528\end{tabular}} &
  \multicolumn{1}{l|}{\begin{tabular}[c]{@{}l@{}}8.193267 \\-j 40.863838\end{tabular}} \\ \hline
\label{tab:tablo2}
\end{tabular}
\end{adjustbox}
\end{table}

\subsection{Generating Dataset}
In the training process of the neural network, we must first obtain the dataset.
It is determined how to create the dataset according to the dataset features. In this study, dataset is created by randomizing the given P and Q values for PQ buses. 
Features to train the ANN are selected based on their influence and ability to change the results of power flow. The random library of Python is used to generate the dataset as random. Load values of the PQ buses are varied within the range of [0.8-1.2]. Then the produced values are fed into a power flow algorithm to solve the power flow problem via Newton Raphson method according to algorithm given in section 2
With using Scikit Learn, data is split to test and training.
The values that are chosen inputs for the training of the neural network are P and Q for PQ buses and V for PV and Slack buses. And the outputs are V for PQ buses and $\delta$ for PQ and PV buses. 

Scaling is one of the most important elements of training a neural network. To determine the scaling technique to be used, other hyperparameters of the neural network are kept constant. 
\section{Simulation and Results}
As mentioned before a new hyperparameter $\beta$ is introduced. It is worth noting that variations of hyperbolic tangent activation function already exist ~\cite{marra2006,gupta2017}. With the constant specifications mentioned in Table~3 simulations for different spin numbers; therefore, for different $\beta$ values were executed. In the study, the training process was completed with multiple $\beta$ values and the most efficient result was obtained with $\beta=4.1$.

\begin{table}[ht]
\centering
\begin{adjustbox}{width=0.45\textwidth}
\begin{tabular}{ll}
\multicolumn{2}{l}{Table 3: Constant Hyperparaeters}                                       \\ \hline
\multicolumn{1}{|l|}{Number of Nodes in Hidden Layer} & \multicolumn{1}{l|}{10}            \\ \hline
\multicolumn{1}{|l|}{Number of Hidden Layer}          & \multicolumn{1}{l|}{7}             \\ \hline
\multicolumn{1}{|l|}{Number of Samples}               & \multicolumn{1}{l|}{3000}          \\ \hline
\multicolumn{1}{|l|}{Batch Number}                    & \multicolumn{1}{l|}{50}            \\ \hline
\multicolumn{1}{|l|}{Epoch}                           & \multicolumn{1}{l|}{50}            \\ \hline
\multicolumn{1}{|l|}{Weight Initializer}              & \multicolumn{1}{l|}{GlorotUniform} \\ \hline
\multicolumn{1}{|l|}{Optimizer Adam}                  & \multicolumn{1}{l|}{Adam}          \\ \hline
\multicolumn{1}{|l|}{Cost Functions Mean Squared Error} & \multicolumn{1}{l|}{Mean Squared Error} \\ \hline
\multicolumn{1}{|l|}{Spin}                            & \multicolumn{1}{l|}{5/2}           \\ \hline
\label{tab:tablo3}
\end{tabular}
\end{adjustbox}
\end{table}

\begin{figure}[ht]%[htbp]
\centerline{\includegraphics[width=3.0in]{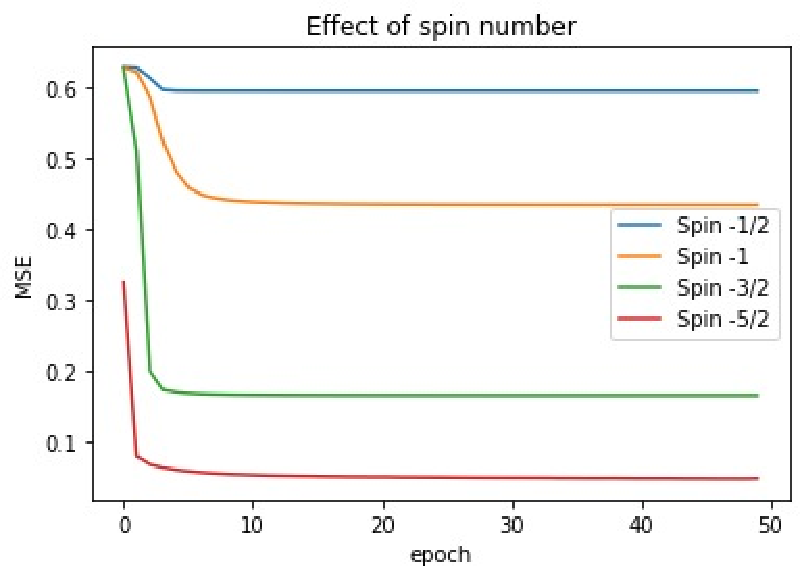}}
\caption{MSE graph in the training process with different spin numbers.}
\label{fig:fig4}
\end{figure}

Hyperparameter specifications of the system are given in Table~\ref{tab:tablo3}.

As can be clearly seen in Fig.~\ref{fig:fig4}, for various beta values according to spin numbers, MSE graphs that change during the learning process were obtained. As can be seen from the graph, the lowest MSE value was obtained for the 5/2 spin number which is refers to $\beta=4.1$.
\subsection{Hyperparameter Tuning}
Through experimentation it has been found that three optimizers prevail for the problem at hand: Adamax, Nadam and Adam. It is also worth noting that to obtain better results weight regularization techniques are introduced. L1-L2 regularization is used with varying hyperparameters for different optimizers. 
It has been observed that mean squared error doubles when number of neurons in the hidden layer drops from 100 to 50 revealing a bottleneck for Adam optimizer. In the end, best solution is obtained with the Adamax optimizer with the specifications given in Table~4.

\begin{table}[ht]
\centering
\begin{adjustbox}{width=0.45\textwidth}
\begin{tabular}{ll}
\multicolumn{2}{l}{Table 4: Adamax Optimal Hyperparameters}                        \\ \hline
\multicolumn{1}{|l|}{Number of Hidden Layer Neurons} & \multicolumn{1}{l|}{50}     \\ \hline
\multicolumn{1}{|l|}{Number of Input Layer Neurons}  & \multicolumn{1}{l|}{10}     \\ \hline
\multicolumn{1}{|l|}{Number of Hidden Layers}        & \multicolumn{1}{l|}{10}     \\ \hline
\multicolumn{1}{|l|}{Batch Number}                   & \multicolumn{1}{l|}{50}     \\ \hline
\multicolumn{1}{|l|}{Epoch Number}                   & \multicolumn{1}{l|}{600}    \\ \hline
\multicolumn{1}{|l|}{L1}                             & \multicolumn{1}{l|}{0.0001} \\ \hline
\multicolumn{1}{|l|}{L2}                             & \multicolumn{1}{l|}{0.0001} \\ \hline
\label{tab:tab4}
\end{tabular}
\end{adjustbox}
\end{table}

It is also mentioned that even though physical realizations are yet to exist, for higher $\beta$ values the ANN performed much better. 
\begin{figure}[ht]%[htbp]
\centerline{\includegraphics[width=3.5in]{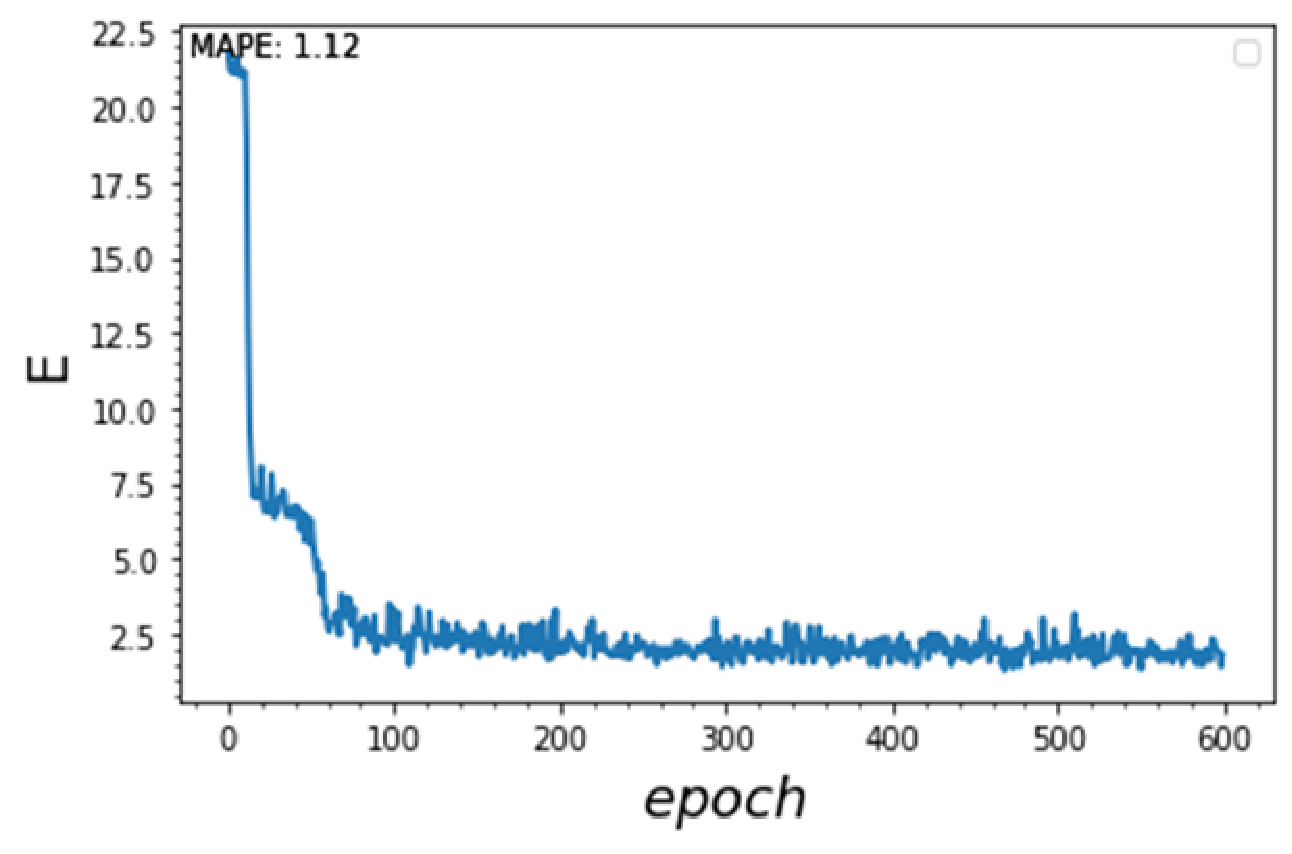}}
\caption{Adamax Optimizer ANN MAPE with $\beta=8$}
\label{fig:fig5}
\end{figure}

\section{Conclusions}
In conclusion, we explore the hypothesis that quantum reservoirs can serve as information sources. We propose an open quantum network model designed to address a specific engineering problem. The proposed hardware consists of open perceptrons operating at the steady state, with spins characterized by $J\geq 1/2$. Following the introduction of the power flow problem, we generate relevant parameters and training sets for the quantum network. Through experimentation, we find that $J\geq 5/2$ is the optimal parameter for minimizing the mean squared error of the quantum network for the given problem.

\section*{Acknowledgment}

The authors gratefully accept funding from the TÜBİTAK (Grant No. 120F353). The authors would also like to thank the Cognitive Systems Lab in the Department of Electrical Engineering for creating a conducive environment for motivating and engaging talks.

\balance

\bibliographystyle{IEEEtran}
%\bibliography{references}


\end{document}